\documentclass[amsmath, aps]{revtex4}
\usepackage{graphicx}
\usepackage{times}
\usepackage{epsfig}
\usepackage{multirow}
\usepackage{ulem}
\begin{document}

\title{Positron production in collision of heavy nuclei}
\author{I.B. Khriplovich\footnote{khios231@mail.ru}, D.A. Solovyev\footnote{solovyev.d@gmail.com}}
\affiliation{ Department of Physics, St. Petersburg State University, Ulianovskaya 1, Petrodvorets, 198504 St. Petersburg, Russia}
\date{\today}

\begin{abstract}
We consider the electromagnetic production of positron in collision of heavy nuclei, with the simultaneously produced electron captured by one of the nuclei. This cross-section exceeds essentially the cross-section of $e^+ e^-$ production.
\end{abstract}

\maketitle

Positron production in collision of heavy nuclei (as well as the production of $e^+ e^-$ pair), was addressed in numerous papers (see, for instance, \cite{mu}-\cite{ma}). In the present note we reconsider the problem of the positron production in collision of heavy nuclei, and derive a simple estimate for the corresponding cross-section.

The diagram related to our problem is presented in Fig. 1. The bold lines in this diagram correspond to the propagation of non-relativistic heavy nuclei with initial momenta $\pm \bar{p}$, and final momenta $\bar{p}_{1,2}$. For the velocities of nuclei we assume $v = 0.1$  (or $v/c = 0.1$ in usual units). The wavy line in Fig. 1 refers to the virtual photon producing the pair $e^+ e^-$.  The produced electron is captured immediately by one of the nuclei, thus creating a single-electron ion; this ion by itself is also on mass shell. The four-momentum of the produced positron is $k_\mu = (k_0, \bar{k})$, $k_0 = \sqrt{k^2+m^2}$, we will neglect here the interaction of positron with nuclei. 
\begin{figure}[hbtp]
	\centering
	\includegraphics[scale=0.5]{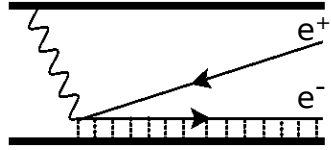}
	\caption{Feynman diagram of positron production. }
\end{figure}

We start the analysis with rather obvious conservation law formulated as follows:
\begin{eqnarray}
\int d\bar{p}_1 d\bar{p}_2 \frac{1}{({\bar{p} -\bar{p}_1)}^4}
\delta\left(\bar{p}_1 + \bar{p}_2 + \bar{k}\right) 
\delta\left(\bar{p}^2/M - \bar{p}_1^2/2M - \bar{p}_2^2/2M - k_0\right).
\end{eqnarray}
Here and below $M$ is the mass of colliding nuclei, and one nucleus has captured the outgoing electron. The factor $1/(\bar{p} -\bar{p}_1)^4$ corresponds to the momentum transfer via the photon exchange between nuclei. Integral (1) is the only expression in this problem which depends on $\bar{p}_1$ and $\bar{p}_2$. Therefore, we can integrate (1) freely over $\bar{p}_1$ and $\bar{p}_2$.

Of course, the nucleus which captures the electron will be slightly more heavy than the another one. We neglect here this tiny difference. Now, integrating (1) over $\bar{p}_2$, we arrive at
\begin{eqnarray}
\label{2}
\int d\bar{p}_1 \frac{1}{\left(\bar{p} -\bar{p}_1\right)^4} \delta(\bar{p}^2/M\,-\bar{p_1}^2/M\, - k_0) ;
\end{eqnarray}
here we have omitted negligibly small term $\bar{p}_1\bar{k}/M$ in the argument of delta-function.

Integral (\ref{2}) is conveniently rewritten as follows:
\begin{eqnarray}
\int \frac{d\bar{q}}{\bar q^4} \delta\left(2\bar{v}{\bar{q}}-k_0\right),
\end{eqnarray}
we have neglected here one more small term $-\bar{q}^2/M$. Let us split now vector $\bar{q}$ into the part parallel to $\bar{v}$, and the part orthogonal to $\bar{v}$, i.e. into $q_{\parallel}$ and  $\bar{q}_\perp$. Then
\begin{eqnarray}
\int\limits^{\infty}_0 \frac{2\pi q_\perp d q_\perp}{\left(\bar q_\perp^2 + q_\parallel^2\right)^2} = \frac{\pi}{q_\parallel^2}.
\end{eqnarray}
At last, we integrate over $q_\parallel$:
\begin{eqnarray}
\pi \int\limits^{\infty}_{-\infty} \frac{dq_\parallel}{q_\parallel^2}\delta\left(2v q_\parallel  -  k_0\right) =\pi \int\limits^{\infty}_{-\infty} \frac{dx}{x^2}
2v \delta(x - k_0) = 2\pi v/k_0^2.
\end{eqnarray}

Let us address now the fermion loop arising in the amplitude squared, see Fig. 2. The upper part of this loop, corresponding to the positron propagation, is $\hat{k}-m$ (as mentioned, we neglect the positron rescattering). The lower part of this loop is a single-electron ion of small velocity $v\sim 0.1$, and we will neglect its velocity at all. Then, the only structure surviving here is
\begin{eqnarray}
Sp\, (\hat{k}-m)\gamma_3\gamma_0\gamma_3 = 4k_0.
\end{eqnarray}
We assume here that the nuclei propagate along the $z$ axis; this is the origin of $\gamma_3$ in the above formula.

The net result of formulas (1) -- (6) is 
\begin{eqnarray}
\frac{2\pi v}{k_0^2}4k_0 = \frac{8\pi v}{k_0}.
\end{eqnarray}
\begin{figure}[hbtp]
\centering
\includegraphics[scale=1.5]{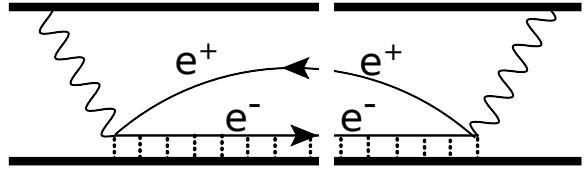}
\caption{The loop, corresponding to positron propagation.}

\end{figure}

Less obvious is the dependence of the discussed cross-section on the velocity $v$ of the colliding nuclei. The common electromagnetic vertex of heavy nucleus is proportional to the velocity $v$, which gives rise to $v^2$ in the cross-section (see Fig. 2). Then, the flux density of nuclei is proportional to their velocity, which gives rise to $v$ in the denominator. Quite non-trivial contribution $\sim v$ follows from (5). Thus, the total cross-section of the process is proportional to $v^2$. 

We consider now in more detail the electron captured by nucleus. The lower part of the fermion loop in Fig. 2 is described by the wave function of bound electron. For the wave function squared of this electron in the momentum representation, we confine to the non-relativistic approximation. This approximation is quite common in such problems, and usually works with a reasonable accuracy. In the present case, this wave function squared is in the momentum representation as follows \cite{be}:
\begin{eqnarray}
\frac{32 \zeta^5}{\pi^2} \frac{1}{\left(k^2+\zeta^2\right)^4};
\end{eqnarray}
here and below $\zeta=Z\alpha m$. 

The cross-section should be multiplied by the factor $2(Z\alpha)^4$. The factor 2 arises here since the electron can be captured either by the upper nucleus, or by the lower one; factor $(Z\alpha)^4$ corresponds to the photon exchange between two nuclei. Then, we should integrate the cross-section over the phase space of positron $\frac{d^3k}{(2\pi)^3(2k_0)}$. In this way, we arrive at the following expression for the cross-section discussed: 
\begin{eqnarray}
\sigma = \frac{256}{\pi} Z^4 \alpha^4 v^2
\int \frac{d^3k}{(2\pi)^3k_0^2}\frac{\zeta^5}{\left(k^2+\zeta^2\right)^4} =
\frac{128 Z^4 \alpha^4}{\pi^3} v^2 \int\limits_0^\infty dk\frac{\zeta^5}{\left(k^2+\zeta^2\right)^4}\frac{k^2}{k^2+m^2}.
\end{eqnarray}
The last integral in this expression is
\begin{eqnarray}
\int\limits_0^\infty dk\frac{\zeta^5}{\left(k^2 + \zeta^2\right)^4}\frac{k^2}{k^2+m^2} = \frac{\pi}{32}\frac{1}{m^2(1+Z \alpha)^4}\left(1+4 Z\alpha +5Z^2\alpha^2\right).
\end{eqnarray}

Thus, the cross-section of positron production in the collision of slow heavy nuclei is
\begin{eqnarray}
\sigma = \frac{128}{\pi^3}\, Z^4 \alpha^4 v^2 \times \frac{\pi}{32}\frac{1}{m^2(1+Z \alpha)^4}\left(1+4 Z\alpha+5Z^2\alpha^2\right) = \frac{4}{\pi^2}\frac{Z^4 \alpha^4}{(1+Z \alpha)^4}(1+4 Z\alpha +5Z^2\alpha^2)\frac{v^2}{m^2}.
\end{eqnarray}
The reasonable estimate for uranium nuclei is
\begin{equation}
\sigma \simeq 0.07\frac{v^2}{m^2}.
\end{equation}

In conclusion, let us come back to the absolute value of the discussed cross-section. According to our estimate (12), it is about $3\times 10^{-23}$ cm$^2$ (for $v=0.1$). This estimate exceeds essentially the cross-section of $e^+ e^-$ production in heavy ion collisions, which is about $10^{-25}$ cm$^2$.



\subsection*{Acknowledgements}

We are grateful to L.N. Labzowsky and V.M. Shabaev for useful discussions.

\end{document}